\documentclass{webofc}
\usepackage[varg]{txfonts}   
\usepackage{graphicx} 
\usepackage{url}
\usepackage{enumitem}

\begin{document}
\title{HPC resources for CMS offline computing: an integration and scalability challenge for the Submission Infrastructure}

\author{
\firstname{Antonio} \lastname{Pérez-Calero Yzquierdo}\inst{1,2}\fnsep\thanks{\email{aperez@pic.es}} \and
\firstname{Marco} \lastname{Mascheroni}\inst{3} \and
\firstname{Edita} \lastname{Kizinevic}\inst{4} \and
\firstname{Farrukh Aftab} \lastname{Khan}\inst{5} \and
\firstname{Hyunwoo} \lastname{Kim}\inst{5} \and
\firstname{Maria} \lastname{Acosta Flechas}\inst{5} \and
\firstname{Nikos} \lastname{Tsipinakis}\inst{4} \and
\firstname{Saqib} \lastname{Haleem}\inst{6}~on behalf of the CMS Collaboration.}

\institute{
Centro de Investigaciones Energ\'eticas, Medioambientales y Tecnol\'ogicas (CIEMAT), Madrid, Spain \and
Port d'Informaci\'o Cientifica (PIC), Barcelona, Spain \and
University of California San Diego, La Jolla, CA, USA \and
European Organization for Nuclear Research (CERN), Geneva, Switzerland \and
Fermi National Accelerator Laboratory, Batavia, IL, USA \and
National Centre for Physics, Islamabad, Pakistan
         }

\abstract{%
The computing resource needs of LHC experiments are expected to continue growing significantly during the Run 3 and into the HL-LHC era. The landscape of available resources will also evolve, as High Performance Computing (HPC) and Cloud resources will provide a comparable, or even dominant, fraction of the total compute capacity. The future years present a challenge for the experiments’ resource provisioning models, both in terms of scalability and increasing complexity. The CMS Submission Infrastructure (SI) provisions computing resources for CMS workflows. This infrastructure is built on a set of federated HTCondor pools, currently aggregating 400k CPU cores distributed worldwide and supporting the simultaneous execution of over 200k computing tasks. Incorporating HPC resources into CMS computing represents firstly an integration challenge, as HPC centers are much more diverse compared to Grid sites. Secondly, evolving the present SI, dimensioned to harness the current CMS computing capacity, to reach the resource scales required for the HL-LHC phase, while maintaining global flexibility and efficiency, will represent an additional challenge for the SI. To preventively address future potential scalability limits, the SI team regularly runs tests to explore the maximum reach of our infrastructure. In this note, the integration of HPC resources into CMS offline computing is summarized, the potential concerns for the SI derived from the increased scale of operations are described, and the most recent results of scalability test on the CMS SI are reported.
}
\maketitle
\section{The CMS Submission Infrastructure}
\label{sec:SI}
\begin{figure}[ht]
\begin{center}
\includegraphics[width=10cm]{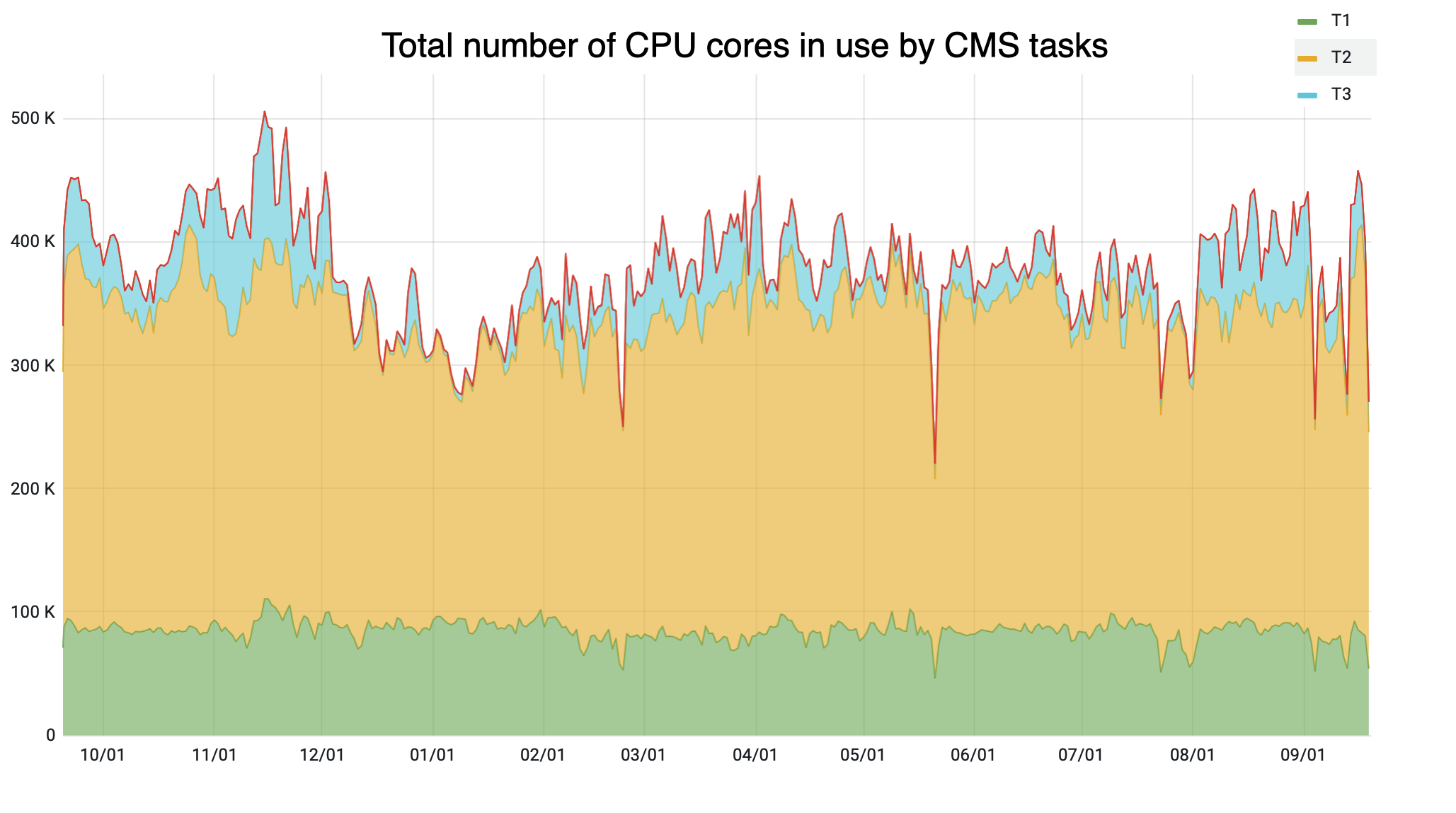}
\caption{Daily averaged number of CPU cores in use by CMS tasks at each of the WLCG Tier layers for the last 12 months.} 
\label{fig:resources}
\end{center}
\end{figure}

The Submission Infrastructure (SI) team in Offline and Computing for the CMS experiment~\cite{cms} is in charge of operating a set of federated HTCondor~\cite{htcondor} pools which aggregates resources from about 70 Grid sites distributed worldwide, plus non-Grid resources, where reconstruction, simulation, and analysis of physics data takes place. In total, the CMS SI manages approximately 400k CPU cores (see Figure~\ref{fig:resources}), regularly supporting the simultaneous execution of nearly 200k CMS computing tasks. 

\begin{figure}[ht]
\begin{center}
\includegraphics[width=10cm]{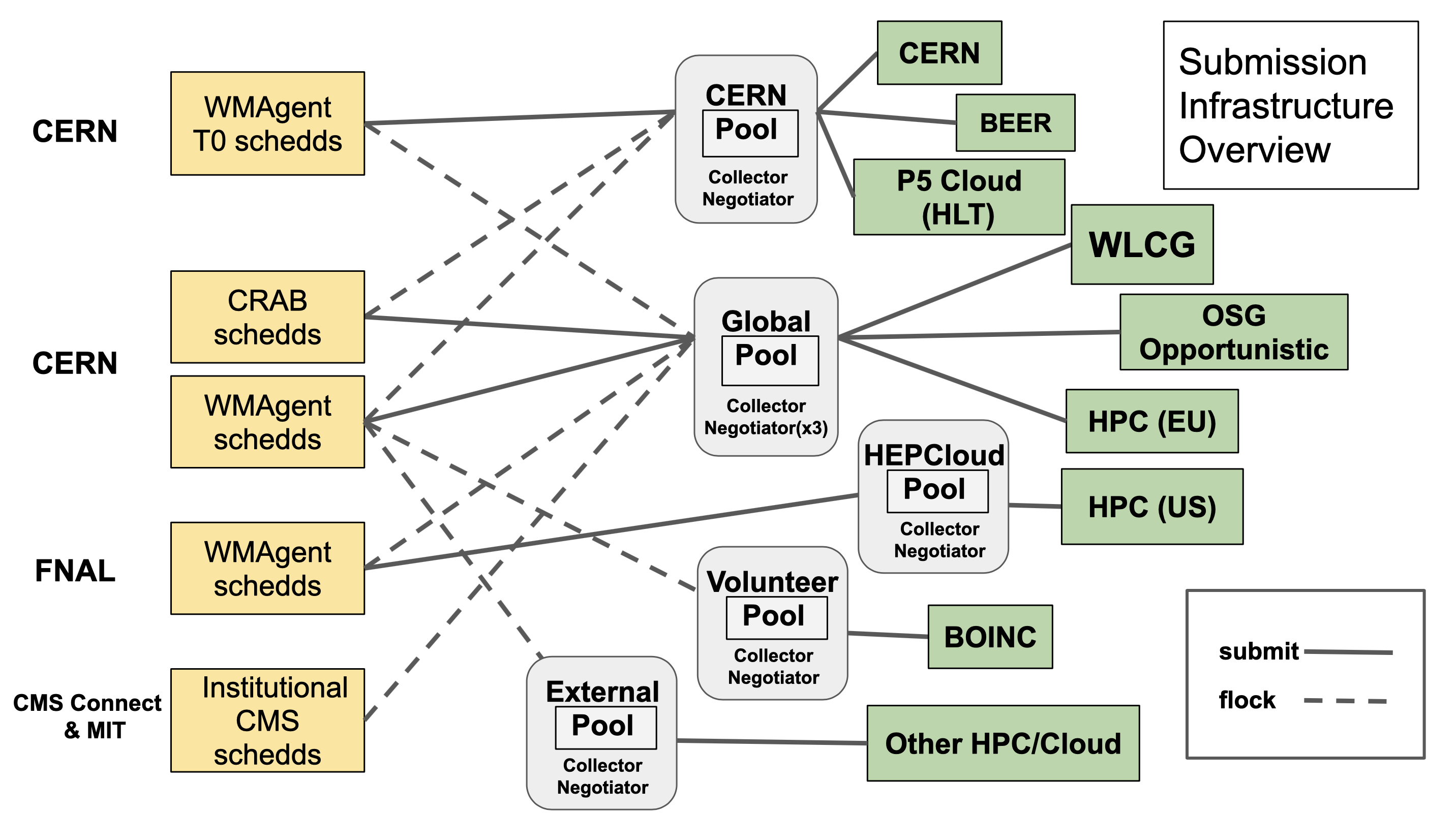}
\caption{CMS SI current configuration, including multiple federated pools of resources aggregated from diverse providers (green boxes) and sets of distributed job schedulers (schedds), handling the Tier-0 and centralized production workloads (WMAgent), as well as analysis job submission (CRAB). A number of opportunistic resources are shown, including Batch on EOS Extra Resources (BEER) at CERN~\cite{beer} and the CMS Run 2 High Level Trigger farm, reconfigured to support CMS offline computing (P5 Cloud~\cite{hlt}).} 
\label{fig:complexity}
\end{center}
\end{figure}

A schematic view of this infrastructure is presented in Figure~\ref{fig:complexity}. Despite its increasing complexity in recent years~\cite{sicomplexity}, its main component continues to be the Global Pool~\cite{globalpool}, which joins the major fraction of resources managed by the CMS SI.

The challenges of operating such an infrastructure reside in managing an ever growing collection of computing resources, connecting new and more diverse resource types (including non-x86 CPU architectures and GPUs) and resource providers, such as WLCG~\cite{wlcg} and OSG~\cite{osg} infrastructures, High Performance Computing (HPC) facilities, as well as Cloud and volunteer clusters, while simultaneously using all of our resources efficiently, maximizing data processing throughput and enforcing task priorities according to CMS scientific program.

\section{HPC growing contribution to CMS computing}
\label{sec:HPC}
HPC centers can help supporting the computing needs of CMS, given the substantial national and supranational investments made and planned for the immediate future. HPC clusters are indeed already part of the scientific computing infrastructure in High Energy Physics (HEP), and it is expected that their contribution will only grow in the coming years. In fact, several examples of HPC resources integration to the CMS computing model have already been completed with success~\cite{CMS_HPC_ACAT}. 

As shown in Figure~\ref{fig:complexity}, CMS follows two main models for the integration of HPC resources to our computing resource portfolio: firstly, in the US, the HEPCloud~\cite{HEPCloud} infrastructure, managed by the FNAL team, supports a separated HTCondor pool, which is however federated to the central CMS SI, being delegated the execution of CMS tasks~\cite{NERSC}. Secondly, a method of WLCG site-extension (transparent from the CMS perspective) is mainly employed for the EU HPC facilities~\cite{KIT}~\cite{bsc}. In this second model, GlideinWMS~\cite{gwms} pilot jobs~\cite{pilots} submitted by the CMS SI to WLCG sites' Compute Elements~\cite{Bird} (CE), are locally redirected onto the corresponding access point of the associated HPC cluster.

\begin{figure}
\begin{center}
\includegraphics[width=10cm]{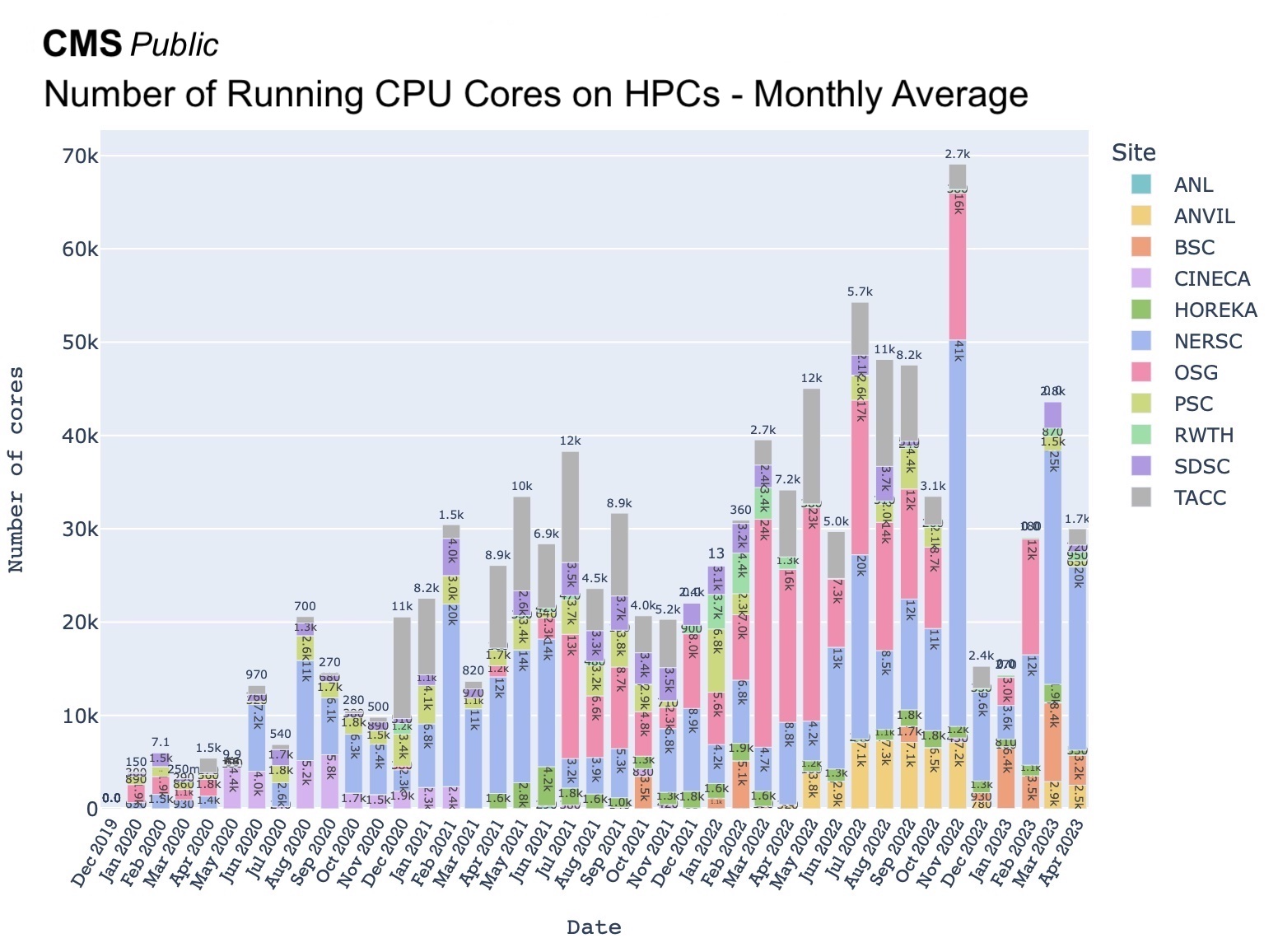}
\end{center}
\caption{HPC contribution to CMS compute resources over the recent years, expressed as a monthly average of CPU cores in use, for each contributing resource provider.}
\label{fig:HPC}
\end{figure}

Figure~\ref{fig:HPC} shows the evolution of HPC resources contribution to CMS offline computing effort over the recent years. A continuous increase in the aggregated CPU usage can be observed, indicating that the compute capacity utilized by CMS at HPC facilities could be sustained at its current level, or even exceeded, in the future years. Moreover, Figure~\ref{fig:HPC} also displays when new facilities are gradually integrated to the CMS HPC resource pool, initially mainly in the USA (e.g. the National Energy Research Scientific Computing center~\cite{hpc_nersc}, NERSC, as well as the Supercomputing Center at the University of San Diego~\cite{hpc_sdsc}, SDSC, and the Pittsburgh Supercomputing Center~\cite{hpc_psc}, PSC). More HPC facilities were then also incorporated to the CMS pool in the EU countries (e.g. with clusters at CINECA~\cite{hpc_cineca} in Italy, HoreKa~\cite{hpc_horeka} in Germany and the Barcelona Supercomputing Center~\cite{hpc_bsc}, BSC, in Spain). 

\section{Challenges in the integration and use of HPC resources by CMS}
Incorporating any new HPC facility to our SI, in order to enable CMS making use of its resources, presents a number of challenges. First, as each HPC facility is different, CMS has to manage a diversity of technical requirements and policy constraints. This affects how access to compute resources, transfers of datasets from/to CMS distributed storage infrastructure, deployment of required software (e.g. CVMFS~\cite{cvmfs}) and services, such as access to remote databases, etc, are technically solved. The integration of HPC clusters thus frequently results in each of these elements having to be resolved case by case. Integration efforts are coordinated centrally by CMS Computing, with the support of the national teams negotiating the use of their respective HPC clusters for CMS offline computing.

\begin{figure}
\begin{center}
\includegraphics[width=10cm]{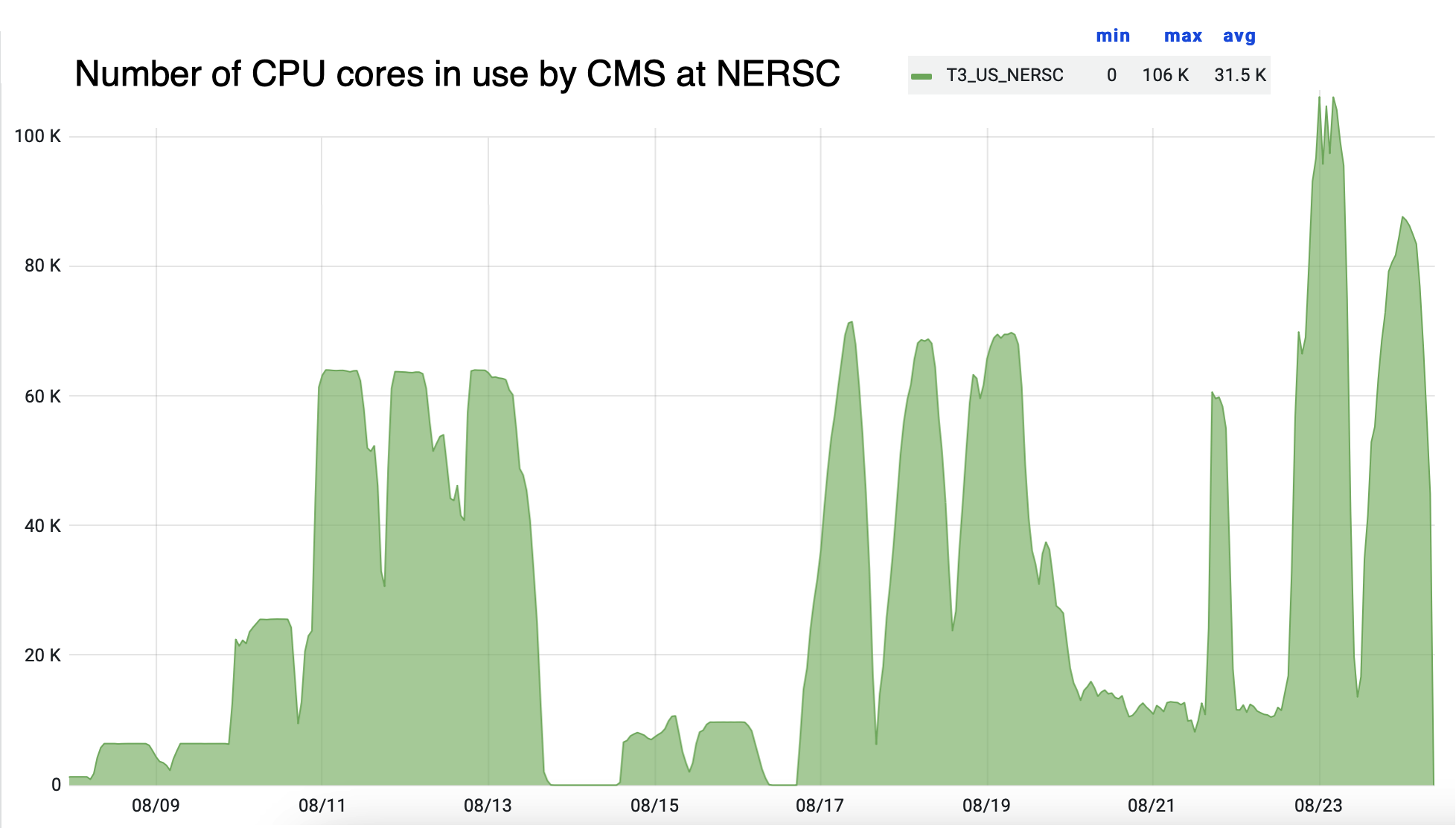}
\end{center}
\caption{Number of CPU cores allocated at NERSC for CMS jobs execution as a function of time, demonstrating a pattern of resource provisioning in peaks usually found in relation to HPC facilities.}
\label{fig:NERSC}
\end{figure}

The second challenge arises from the scale of potentially available resources at HPCs, and the fact that CPU allocation at these facilities typically differs from resource provisioning at WLCG sites. Indeed, under sufficient workload pressure from CMS, grid sites tend to supply a constant amount of CPU power, generally in agreement with the capacity each site has pledged to the experiment. On the contrary, HPC facilities often allocate CPU resources to their users in shorter yet more intense bursts. An example of this pattern, at NERSC, is shown in Figure~\ref{fig:NERSC}, including peaks of up to 100k CPU cores in use by CMS lasting for a few hours of duration.

The CMS Workload Management (WM) and SI systems must therefore be appropriately dimensioned to manage and distribute workloads in order to leverage these enlarged pools of resources, profiting from the HPC resource peaks in addition to the WLCG resource baseline. Ensuring the stability and flexibility of the WM and SI infrastructures and services is crucial for effectively harnessing the increased compute capacity provided by HPCs, which will be discussed in the following sections.

\section{Scalability limits in the CMS SI}
\label{sec:scalability}
Operating away from any scalability limiting factor is a critical aspect for a system such as the CMS SI, designed to perform in a dynamic environment, adapting itself to the growing CMS resource demands, the fluctuating resource availability from the WLCG and other providers, and the variable mix of workloads the SI has to manage. The SI team has therefore proactively run scalability tests over the years in order to find those limits for every dimension of our system, and, consequently, evolve the infrastructure to push the scalability limits further away.

The assessment of SI scalability involves firstly analyzing the computing capacity that the Central Manager (CM) element of our HTCondor pools can effectively harness. This is in turn derived from the $collector$ capability of real-time processing the stream of status updates arriving from the slots across the pool, as well as keeping the $negotiator$ matchmaking cycle time under control. Secondly, once the resources are available and can be effectively negotiated, the active pool of scheduler nodes ($schedds$, see Figure~\ref{fig:complexity}) must be tested on their aggregated maximum concurrent job execution capacity. This metric determines the number of workflows and the scale of resources that can be efficiently managed by our infrastructure. 

\subsection{Pushing the limits of the CMS Global Pool}
The SI team is continuously working on detecting and solving scaling bottlenecks to the infrastructure, with the support of the HTCondor and GlideinWMS developers teams. As a consequence of the accumulated experience, our setup currently includes multiple "non-standard" customized settings, most of them aimed at avoiding the saturation of the main collector service of the pool, as described in previous reports~\cite{siscalability}. Some of these specialized settings are: 

\begin{enumerate}[label=(\alph*)]
    \item Our HTCondor Connection Broker (CCB) service is running on a separate host to the CM, and configured with an enlarged pool of available connection sockets.
    \item Our Global Pool CM employs multiple negotiator daemons, running in multi-threaded mode, to speed up the negotiation cycle.
    \item Multiple secondary collector processes are deployed in our CM, supporting the task of the main top collector daemon.
    \item The stream of slot update messages to the collectors has been optimized to only propagate certain status transitions (such as the slot becoming $unclaimed$ by any user, thus subject to negotiation in the following cycle), filtering out the rest. 
    \item Slot update messages use UDP protocol instead of TCP (also the CM is configured with an enlarged UDP buffer size).
    \item Pool status queries reaching the collector from the negotiator are classified as high-priority, as opposed to those launched by other services in the SI, and from the monitoring processes. Non high-priority queries are subsequently redirected to a secondary collector for the pool (in fact, the backup infrastructure running at FNAL, see~\cite{sistability})
\end{enumerate}


\subsection{The SI 2023 scale tests}
\label{sec:tests}
The main goal for the Spring 2023 tests of the CMS SI was to assess the potential scalability of our Global Pool, considering the following updates since our latest tests (2021, see~\cite{siscalability}). These include the evolution in HTCondor software (tested version 10.0.1), a new physical host for the CM processes (AMD EPYC 7302 at 3 GHz), the adoption of token-based authentication for HTCondor services in our SI~\cite{tokens}, and the incrementally improved configuration of our infrastructure. As in previous tests, $uberglideins$~\cite{uberglideins}, running multiple HTCondor $startd$ processes per glidein, are used to generate enlarged copies of our Global Pool. A secondary collector daemon is employed as a source of monitoring data for the test pool, reducing the stress from non-essential queries on the main collector.

Initial results indicated a first bottleneck was found in the total capacity of our pool of schedds to supply jobs to the resource pool. At 1 MB of RAM consumed per running job, our schedds, hosted on VMs configured with 50 GB of memory, saturated at approximately 50k running jobs each. With an initial set of 10 identical schedds, the saturation of our test infrastructure was observed at 500k simultaneously running jobs, comparable to our previous result. The set of schedds was therefore enlarged with new machines, such that the new aggregated capacity was, in principle, capable of supporting approximately 1M running jobs. 

However, a second bottleneck immediately appeared in the following tests, caused by the CCB service, initially hosted in a VM deployed over a shared hypervisor, which was effectively limiting the maximum number of TCP connections that the broker could manage. As a consequence, open slots could not be matched, therefore constraining the pool from reaching higher scales. The CCB service was then redeployed to a physical node under SI team responsibility, which could then optimize its network configuration, opening again the possibility of reaching higher pool scales. 

\begin{figure}
\begin{center}
\includegraphics[width=12cm]{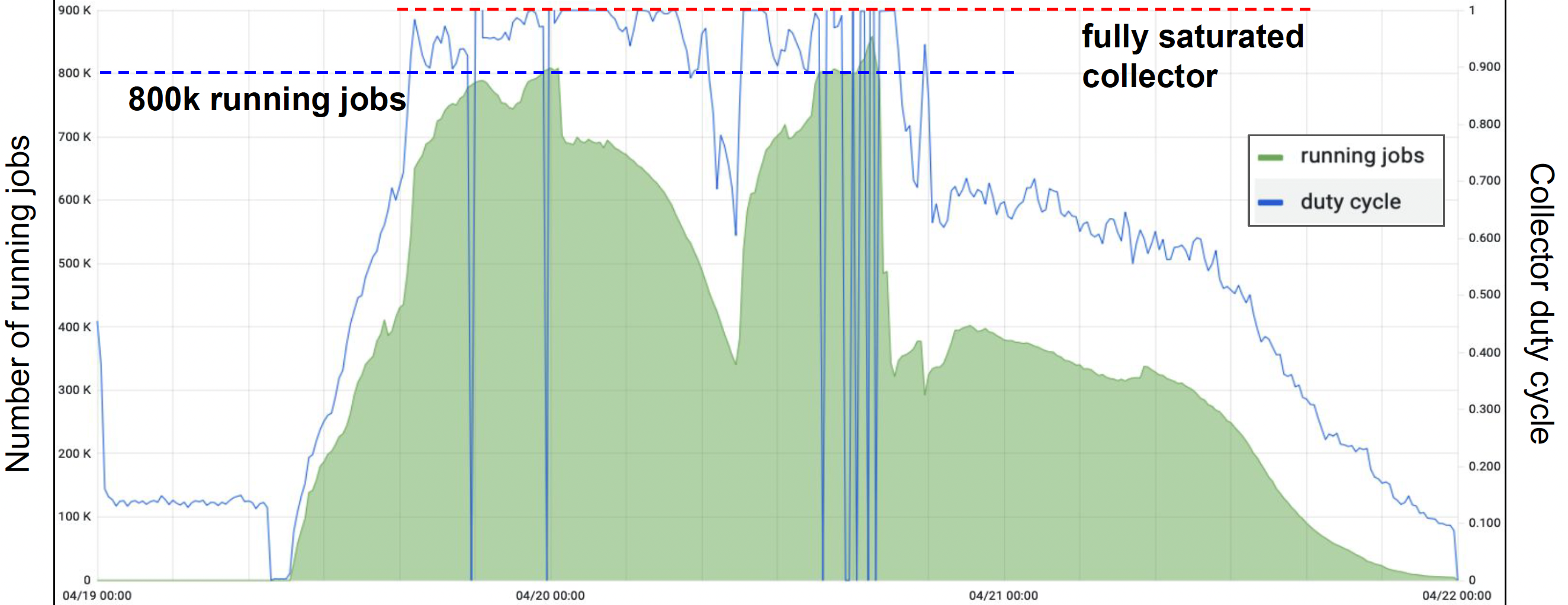}
\end{center}
\caption{Final results of the Spring 2023 scalability tests of the CMS HTCondor Global Pool, reaching the simultaneous execution of approximately 800k jobs. The collector duty cycle indicates the fraction of seconds over a predefined time interval that the collector process was busy, i.e. not idle waiting to process incoming requests.}
\label{fig:2023_scale}
\end{figure}

Ultimately, the saturation of the primary pool collector set the boundary for further scaling of our infrastructure, in a sequence that can be summarized as follows: the collector, which processes slot status updates, becomes increasingly stressed as the size of the pool, and thus the rate of updates, keeps growing. When the collector daemon becomes fully saturated (indicated in Figure~\ref{fig:2023_scale} by collector duty cycle approaching 1), it can’t provide further updated slots status information to the negotiator. Jobs to slots matchmaking process becomes therefore inefficient, leaving many slots unused. This effectively limits the total number of running jobs that our test infrastructure can manage while still using the pool resources efficiently. As Figure~\ref{fig:2023_scale} shows, our latest tests pushed the scalability of our Global Pool to about 800k simultaneously running jobs. This represents a potential for a factor 4 growth in comparison to our current number of running jobs.

\section{Conclusions and future work}\label{sec:conclusions}
The contribution of HPC resources to CMS Computing has increased in the recent past and is expected to continue growing in the coming years. The integration of these resources presents a number of challenges to our infrastructure, including that of scalability, caused by the increasing total amount of compute resources, but more importantly, due to the frequently observed pattern of CPU allocation at HPCs, in bursts.  

The CMS SI team, in collaboration with HTCondor and GlideinWMS developers, performs regular assessments of infrastructure scalability. This proactive approach aims to anticipate future scaling challenges and ensures that we remain ahead of any potential limitations. Our Spring 2023 scalability tests demonstrated the capacity of our SI to support up to 800k simultaneous running jobs, which indicates a potential for fourfold growth compared to CMS current number of running jobs. The LHC program extends well into the future, therefore continuously pushing the SI for higher scales, as required by CMS needs, while maintaining stability and efficiency, is a critical element of the future CMS computing program success.

\section*{Acknowledgements}
This work was partially supported by the Spanish Ministry of Science and Innovation under grants PID2019-110942RB-C21, PID2019-110942RB-C22 and PID2020-113807RA-I00, which include FEDER funds from the European Union, and by the US National Science Foundation under Grant No. 2121686.

\end{document}